# Highly Sensitive Differential Microwave Sensor for Soil Moisture Measurement

Rasool Keshavarz, Justin Lipman, *IEEE Senior Member*, Dominique Schreurs, *IEEE Fellow Member,* and Negin Shariati, *IEEE Member*

***Abstract*— This paper presents a highly sensitive differential soil moisture sensor (DSMS) using a microstrip line loaded with triangular two-turn resonator (T2-SR) and complementary of the rectangular two-turn spiral resonator (CR2-SR), simultaneously. Volumetric Water Content (VWC) or permittivity sensing is conducted by loading the T2-SR side with dielectric samples. Two transmission notches are observed for identical loads relating to T2-SR and CR2-SR. The CR2-SR notch at 4.39 GHz is used as a reference for differential permittivity measurement method. Further, the resonance frequency of T2-SR is measured relative to the reference value. Based on this frequency difference, the permittivity of soil is calculated which is related to the soil VWC. Triangular two-turn resonator (T2-SR) resonance frequency changes from 4 to 2.38 GHz when VWC varies 0% to 30%. The sensor's operation principle is described through circuit model analysis and simulations. To validate the differential sensing concept, prototype of the designed 3-cell DSMS is fabricated and measured. The proposed sensor exhibits frequency shift of 110 MHz for 1% change at the highest soil moisture content (30%) for sandy-type soil. This work proves the differential microwave sensing concept for precision agriculture.**

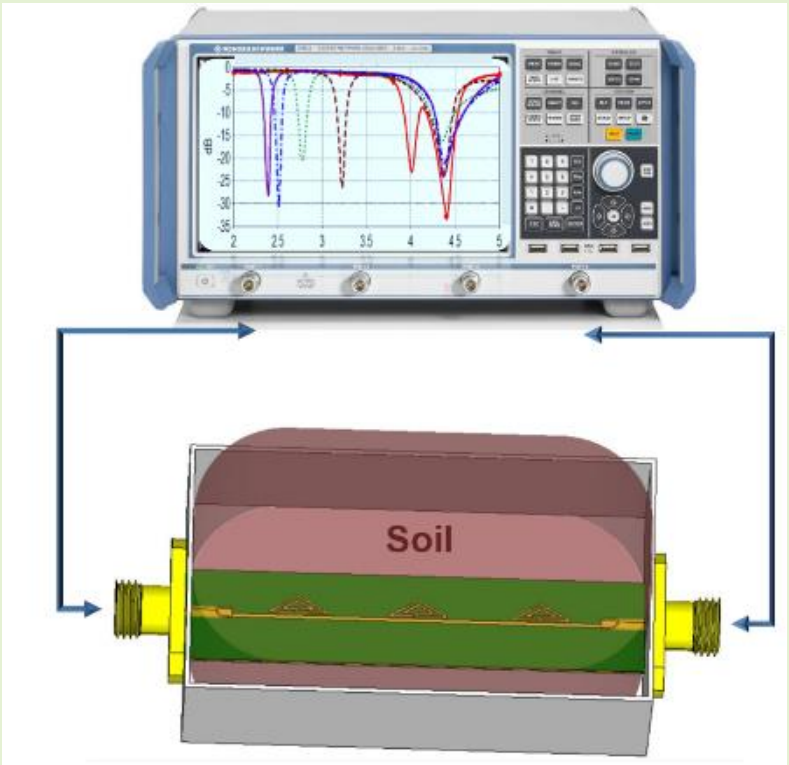

## I. Introduction

EFFICIENT utilization of agricultural resources for improved production and reduced environmental impact is the basis of precision farming. In a conventional agricultural measurement method, a network of sensor nodes distributed over a wide area is used to model intra- and inter-field variations. Each sensor transmits local features of the soil that surrounds it. Using the Internet-of-Things (IoT), collected data is sent to the base station that analyses data and takes necessary steps, like irrigation and fertilization [1], [2]. Nowadays, a broad range of IoT devices is implemented in numerous applications with the expansion of modern farming. Therefore, lightweight, low-cost, high-precision and low power consumption sensors are desirable for precision farming [3], [4], [5]. Moreover, a great advantage of IoT technology is enhancing the accuracy of sensors in the smart farming [6].

As a result of a large discrepancy between the relative dielectric properties (real part) of liquid water (approximately 80 for less than 5GHz) and dry soil (2 to 5), soil moisture content can be determined from the real part of the soil dielectric constant measurements. Due to the Volumetric Water Content (VWC), the imaginary portion of the soil dielectric constant affects the insertion loss in the measurement procedure [7].

Different methods have been proposed to estimate the soil permittivity based on the identified criteria; time-domain reflectometry (TDR), capacitance, frequency domain reflectometry (FDR), remote sensing, etc. [8], [9], [10]. FDR estimates the soil moisture content based on frequency variations of a signal due to the dielectric properties of soil. In recent years, planar FDR microwave sensors have been attracted many researchers due to their simple production, low fabrication cost, high sensitivity, reliability, and design versatility [7]. However, having a low-quality factor is one of the drawbacks of planar microwave sensors that could limit their applications. To address this, some techniques have been presented to produce ultra-high-Q microwave sensors [11], [12]. Another disadvantage of planar frequency domain microwave sensors is their high sensitivity to slight environmental variations such as temperature, soil composition, etc. This requires differential solutions to take such variables into account. Since environmental factors are common-mode parameters, their effect will be removed in differential measurements. In frequency domain sensors, the differential property is achieved by considering two (or more) resonance frequencies, where one of them is acting as the reference [13], [14], [15].

Metamaterials structures [16], [17], especially split-ring resonator (SRR) and its complement (CSRR) have been widely used as sensing devices, due to their high-quality factor and small size [18]. Their applications have been found in biomolecule identification, concentration analysis, and microfluidic characterization. Penetration of electromagnetic

RF and Communication Technologies (RFCT) research laboratory, University of Technology Sydney, Ultimo, NSW 2007, Australia, e-mail: Rasool.Keshavarz@uts.edu.au;

waves into materials in the vicinity of a sensor lays the groundwork for remote sensing [19], [20]. Recently, several shapes of SRR and CSRR have been demonstrated where each of them exhibits different benefits relative to each other [21].

This paper proposes a new low-profile and high-Q differential soil moisture sensor (DSMS) based on a microstrip line which is loaded with two resonators simultaneously; a triangular 2 turn spiral resonator (T2-SR) and a complement of the rectangular spiral resonator (CR2-SR). In the proposed structure, the CR2-SR acts as the reference resonator to eliminate environmental conditions. Moreover, compared to conventional SRRs, a T2-SR with the same electrical size provides a higher Q and a stronger resonance which enhances the sensitivity and accuracy of the proposed DSMS. Due to its compactness, resolution, and low-cost properties, the proposed sensor has a potential to be used as a soil moisture sensor in precision farming.

Contributions of this paper are summarized as follows:

- As a result of combining two uncoupled resonators (T2-SR and CR2-SR) in the sensor structure, the proposed DSMS exhibits differential measurement property to create a highly reliable measurement procedure.
- The proposed highly sensitive sensor has been presented for soil moisture measurement which is related to the permittivity range of 3 to 16.5 (for sandy soil). Further, this methodology is applicable over a wide range of permittivity measurement scenarios. The proposed technique can be adopted to different material detection systems.
- From a design and economical perspective, utilizing both sides of PCB (top and bottom layers) to realize two types of resonators in each layer, results in decreasing the total footprint. Therefore, the proposed differential sensor is compact, low cost and has a potential to be embedded into practical applications.
- An accurate design guide procedure and equivalent circuit model have been presented and equations were derived.

The organization of this paper is as follows: Theory and design principles of the proposed DSMS are presented in section II. The DSMS performance is validated by simulation and measurement results in section III. Finally, conclusions are provided in section IV.

## II. Theory and Design Principle of The Proposed DSMS

The schematic (top and bottom layers) and 3D layers of the proposed DSMS are presented in Fig. 1(a) and 1(b). This sensor consists of three parts: microstrip line, triangular 2-turn spiral resonator (T2-SR), and complementary rectangular 2-turn spiral resonator (CR2-SR). The T2-SR includes 2-turn concentric spiral metallic rings, printed on a substrate (top layer) and is edge-coupled to the microstrip line, while the CR2-SR is etched on the ground plane (bottom layer) (Fig. 1(b)). As the coupling mechanism of T2-SR and CR2-SR are dominated by electrical and magnetic fields, respectively, these two resonators are not coupled to each other and hence, their resonance frequencies are isolated. From a practical perspective, two distinct resonators are implemented to calibrate the environmental impacts. Therefore, the extraneous environmental effect is subtracted using a differential sensing algorithm. Fig. 1(b) exhibits sensor layers in the measurement setup; reference layer (air or foam) and material under test (MUT) layer with permittivity of $\varepsilon_{r0}$ and $\varepsilon_m$, respectively. Between these two layers, the proposed sensor is placed on a microwave substrate with permittivity and thickness of $\varepsilon_r$ and $h$, respectively. The reason of etching T2-SR on the top layer and CR2-SR on the bottom layer as the reference will be explained later in this section.

Due to small electrical dimension of the T2-SR and CR2-SR at resonance frequencies, microstrip loaded lines can be described by equivalent circuits of lumped element. The equivalent circuit model (unit cell) for the T2-SR and CR2-SR loaded transmission line is shown in Fig. 1(c). $L$ and $C$ are per-section inductance and capacitance of the microstrip line. T2-SR is modeled as a resonant tank which is magnetically coupled to the line through mutual inductance, $M$. This produces inductance $L_s$ and capacitance $C_s$ after tank transformation to horizontal branch in the equivalent circuit model (Fig. 1(c)) [22]. Whereas, CR2-SR is mainly excited by the electric field induced through the line. This coupling can be modeled by connecting the series line capacitance to CR2-SR, which is modeled as a parallel $LC$ tank. According to this, the proposed lumped element equivalent circuit model for the CR2-SR loaded transmission line (Fig. 1.c) consists of $L_c$ and $C_c$ as the CR2-SR. Due to the proposed sensor dimensions and for simplicity, the equivalent circuit model is assumed lossless. This assumption only affects Q-factor of the resonances and not the resonance frequencies. Figure 1(d) shows the E-field distribution of the unloaded DSMS at the resonance frequencies for the top and bottom resonators.

Effective permittivity of a microstrip line that is buried under a MUT can be expressed as [23]:

$$\varepsilon_{eff} = \varepsilon_0\left[(\varepsilon_r + \varepsilon_m)/2 + \left((\varepsilon_r - \varepsilon_m)/2\sqrt{1 + 12\frac{h}{W_a}}\right)\right] \quad if\ T \gg h \qquad (1)$$

Where $W_a$ is the average width of microstrip line and $T$ is the MUT thickness.

According to Fig. 1(c), transmission response of the proposed DSMS exhibits two zeros (resonance frequencies) at:

$$f_{z1} = \frac{1}{2\pi\sqrt{L_s C_s}}, f_{z2} = \frac{1}{2\pi\sqrt{L_c(C_c + C)}} \qquad (2)$$

Where $f_{z1}$ and $f_{z2}$ are resonance frequencies of the T2-SR and CR2-SR, respectively.

Fig. 2 shows S-parameters results (full-wave and circuit model) of the proposed 3-cell DSMS ($\varepsilon_m = 1$). This figure confirms accuracy of the proposed equivalent circuit model and design procedure. Values of the equivalent circuit model parameters are presented in Table I. These parameters have been extracted from closed-form equations [23] and optimization process using Advanced Design System (ADS). Moreover, the geometrical dimensions of T2-SR, CR2-SR, and transmission line are provided in Table II.

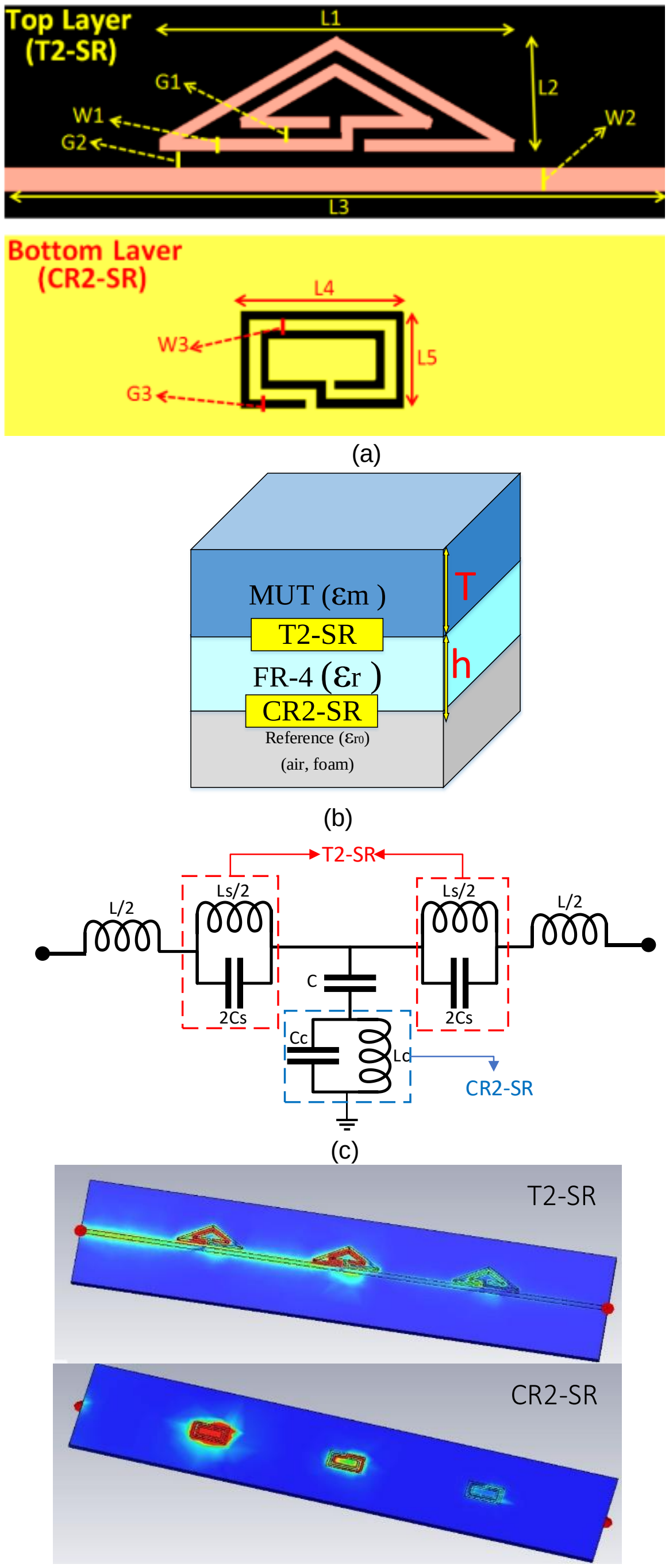

Fig. 1. Proposed DSMS unit cell a) DSMS schematic (top and bottom layers), b) 3D Layers, c) Equivalent circuit model, d) E-field distribution of the proposed DSMS for three cells.

In the proposed sensor, two resonance frequencies are completely uncoupled; one is kept immune from variations known as static/reference resonance (CR2-SR) and the other is used for conventional sensing as a dynamic resonance (T2-SR). The inductors in the sensor equivalent circuit model ($L_s$, $L_c$) are not impacted by the MUT permittivity and they are roughly constant in the measurement process. Therefore, key components in the analysis of the DSMS resonance frequencies are capacitances ($C$, $C_c$, $C_s$). The effect of MUT permittivity on the capacitances values is investigated in this paper.

The capacitance $C_c$ in Fig. 1(c), corresponds to a metallic rectangular sheet that is surrounded by a ground plane at a distance $g$. An analytical and complicated expression for this type of capacitance is derived and presented [23]. However, in this paper, we used a numerical solution to find $C_c$ equivalent capacitance based on this expression. In contrast, finding $C_s$ and $C$ is straightforward and can be calculated as in [23]:

$$C_s = C_{pul} \times P_{se},\ C = \frac{l\sqrt{\varepsilon_{re}}}{cZ_0} \qquad (3)$$

Where $C_{pul}$ is the per unit length capacitance along with the slot between square rings, $P_{se}$ is the effective average perimeter of two triangular rings in the T2-SR structure, $l$ and $Z_0$ are the length and characteristic impedance of the microstrip transmission line, $\varepsilon_{re}$ is the effective permittivity of the structure and c denotes the light velocity.

Now, the sensitivity is defined as the variation of resonance frequency ($f_{z1}$ or $f_{z2}$) relative to the permittivity (or VWC) of the MUT as:

$$S_{fzi} = \frac{\Delta f_{zi}}{\Delta(VWC)} = \frac{\Delta f_{zi}}{\Delta\varepsilon_m} = \frac{\Delta f_{zi}}{\Delta C_{eq}} \times \frac{\Delta C_{eq}}{\Delta\varepsilon_m} = \frac{-1}{4\pi\sqrt{L_c (C_{eq})^3}} \times \frac{\Delta C_{eq}}{\Delta\varepsilon_m} \quad (i = 1,2) \qquad (4)$$

where $C_{eq} = C_s$ and $C_{eq} = C_c + C$ are to calculate $S_{fz1}$ and $S_{fz2}$, respectively. Therefore, considering sensitivity of the equivalent capacitance relative to permittivity ($\Delta C_{eq}/\Delta\varepsilon_m$), $S_{fzi}$ is calculated. Finally, according to (1), (2) and (3) and using ADS, the optimum values of the DSMS components are obtained to maximize sensor sensitivity over a broad range of permittivity at desired frequency bands.

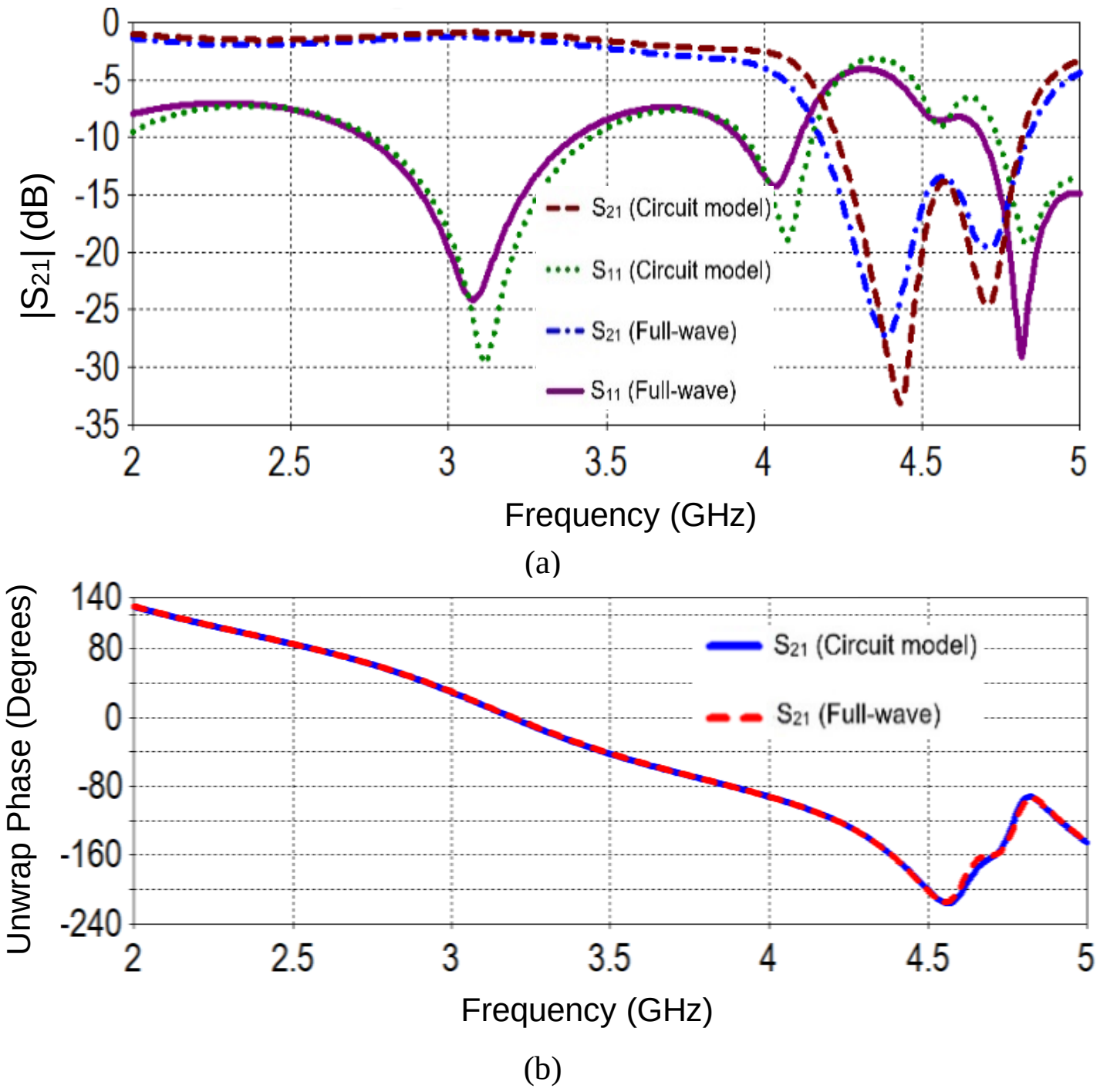

Fig. 2. Full-wave and circuit model simulation results of the proposed 3-cell DSMS ($\varepsilon_m = 1$), a) |S21|, b) unwrap phase(S21).

TABLE I. VALUES OF THE EQUIVALENT CIRCUIT MODEL.

| $C$ (pF) | $L$ (nH) | $L_s$(nH) | $C_s$ (pF) | $L_c$(nH) | $C_c$ (pF) |
|---|---|---|---|---|---|
| 1.73 | 4.6 | 3.85 | 1.11 | 4.33 | 1.84 |

TABLE II. DIMENSIONS OF THE PROPOSED DSMS IN (FIG. 1(A))

| L1 | L2 | L3 | L4 | L5 | W1 | W2 | W3 | G1 | G2 | G3 |
|---|---|---|---|---|---|---|---|---|---|---|
| 6.2 | 1.9 | 16.2 | 2.1 | 3.3 | 0.2 | 0.4 | 0.2 | 0.2 | 0.2 | 0.2 |

## III. Simulation, Measurement, and Discussion

Regarding the number of cells in the proposed DSMS, there is a tradeoff between the sensor resolution (bandwidth of the notch frequencies) and $|S_{21}|$ value (null depth) at two notch frequencies related to T2-SR and CR2-SR. As a case study, Fig. 3(b) presents $|S_{21}|$ simulation result for different numbers of unit cells at $\varepsilon_{rm} = 10$. According to this figure, by increasing the number of cells from 1 to 5, notch bandwidths and null depth increase. Therefore, considering a 3-cell DSMS leads to a good balance between notch bandwidth and null depth.

In order to verify the proposed sensor functionality, the designed 3-cell DSMS is fabricated on an FR4 substrate with a thickness of 0.6 mm, relative permittivity 4.6, and loss tangent 0.01 (Fig. 4). Moreover, a foam box was placed around the sensor to pour soil on the sensor easily and isolate its bottom layer from the soil (Fig. 4(c)). The foam permittivity is around 1.2 which is close to air permittivity. According to measurement results, a foam thickness of 5 mm ($0.1\lambda_g$ at 3.5 GHz) is sufficient to eliminate the effect of soil mass on the buried sensor bottom. Measurements were performed using Vector Network Analyzer (VNA-ZVA40).

As mentioned in section II, the concept of permittivity (or soil VWC) measurement of the proposed DSMS is based on the resonance frequency variation, relative to the MUT permittivity. Several relationships between soil moisture content and soil dielectric constant have been proposed [24], [25]. Table III presents real parts of the dielectric constant values as a function of VWC (0-30%) in S-band (2-4 GHz). The sandy soil is wet above VWC of 30%, and cannot be considered as moist soil.

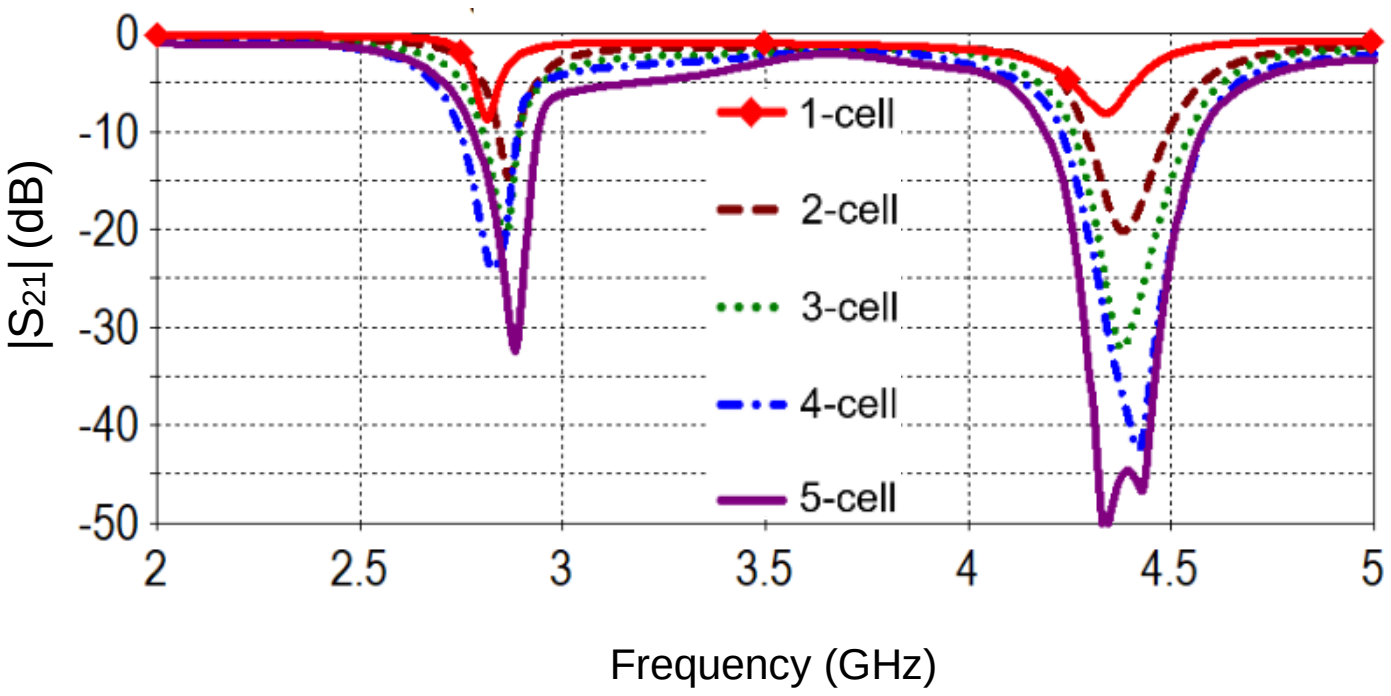

Fig. 3. a) $|S_{21}|$ simulation result for different numbers of unit cells at $\varepsilon_m = 10$.

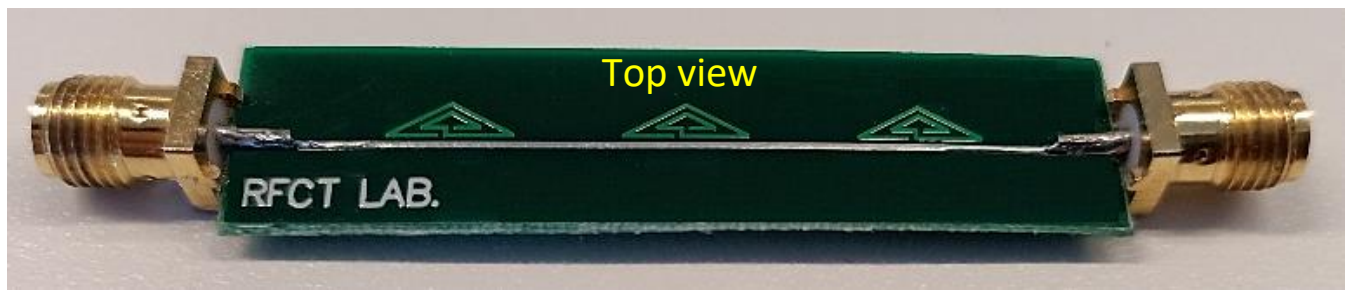

(a)

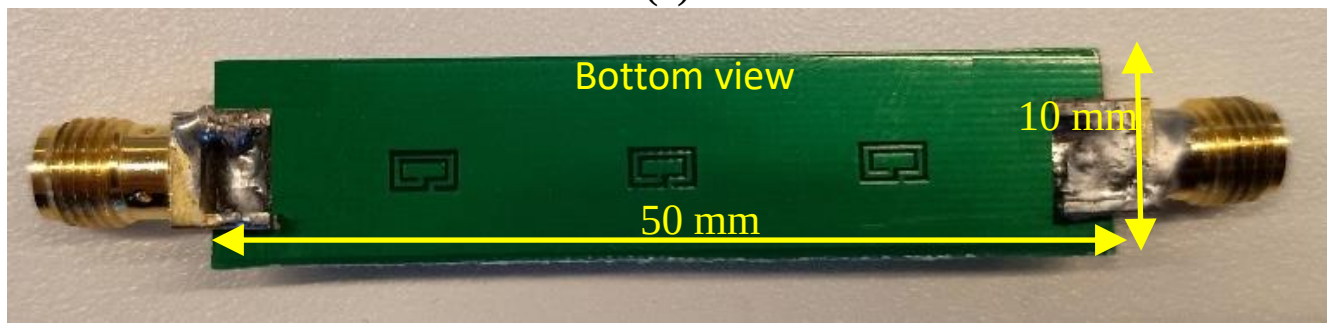

(b)

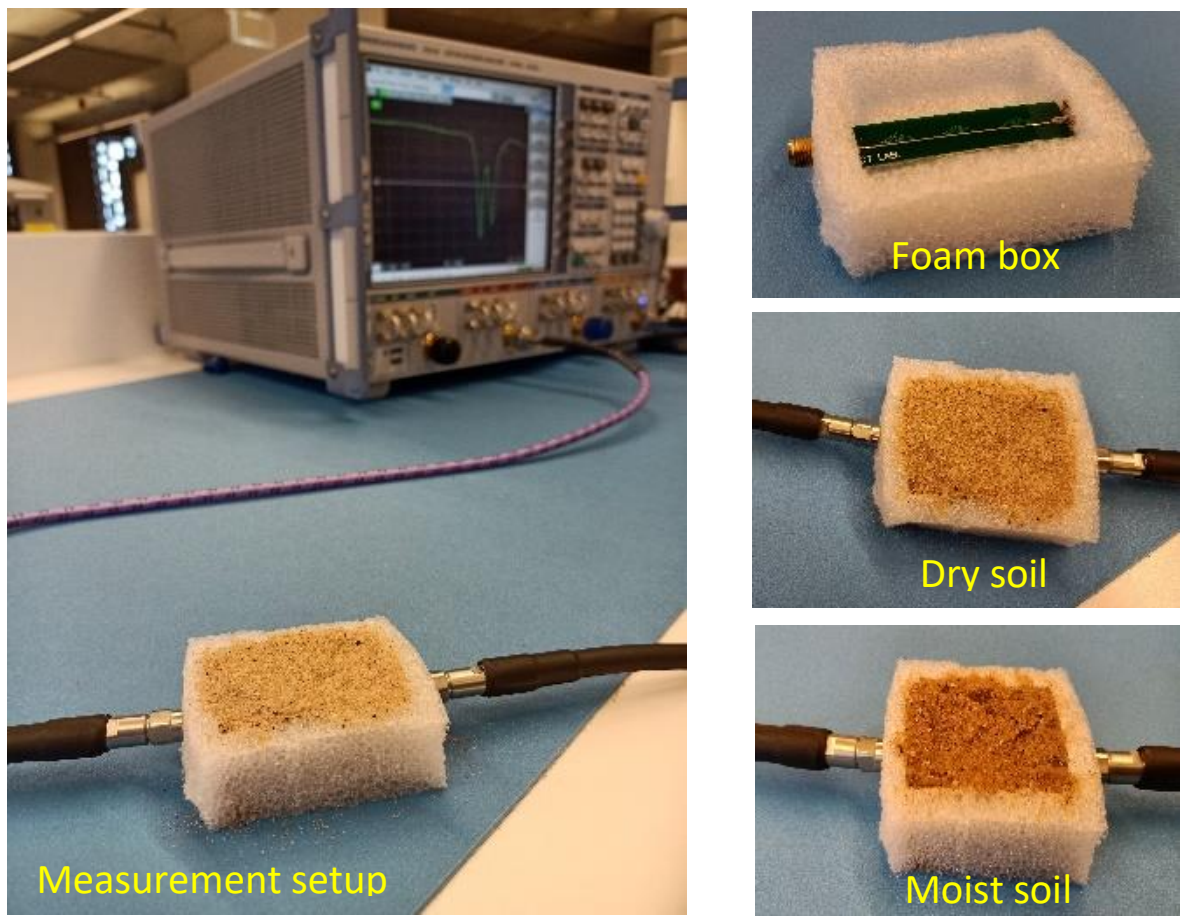

(c)

Fig. 4. Prototype of the proposed 3-cell DSMS consisting of a transmission line loaded with T2-SR and CR2-SR on FR4 substrate with $\varepsilon_r$ = 4.6, thickness 1.6 mm, and loss tangent 0.01 a) top view, b) bottom view, c) measurement setup.

According to Fig. 1(b), the CR2-SR is unloaded ($\varepsilon_{r0} = 1$) as a reference resonator in simulations and the T2-SR is loaded by a 30 mm thick dielectric slab, where the dielectric slab covers the whole resonator area. Permittivity of the dielectric slab ($\varepsilon_m$) is varied in each simulation based on the values in Table III, and the corresponding transmission response (|S21|) is monitored (Fig. 5(a)). Figure 5 shows two transmission zeros for different VWC values, related to T2-SR and CR2-SR resonators which are presented in (2). These two frequencies are very close for VWC=0%, and with increasing VWC to 30%, $f_{z1}$ decreases where the reference transmission zero ($f_{z2}$) is constant for different values of $\varepsilon_m$ (or VWC). Hence, this property can be used in differential sensing.

Moreover, after validating the proposed sensor performance in unloaded state, the structure has been embedded into the sandy soil with different VWC. Different soil VWC levels have been achieved based on [25] that has been released by the Department of Sustainable Natural Resources NSW Australia. Firstly, the sand has been put in the oven to dry it. Then, the water has been added to provide sand with 10%, 20%, 25%, and 30% VWC levels as:

$$VWC(\%) = \frac{W_2}{W_1+W_2} \times 100 \quad (5)$$

Where $W_1$ and $W_2$ are the weights of dried soil and added water, respectively. Furthermore, we measured the performance of proposed sensor in 5 trials for each VWC value and then presented average values in Fig. 6(b). According to Table III, for the soil moisture value range of 0% to 30%, $\varepsilon'_m$ varies from 3.7 to 16.7 and the measured resonance frequency of T2-SR changes from 4 GHz to 2.38 GHz, while the resonance frequency of CR2-SR is almost constant around 4.39 GHz in measurements (Fig. 5(b)). The measurement result verifies that the T2-SR and CR2-SR are uncoupled in the designed sensor.

Table III. Dielectric constant for sand at S-band (2-4 GHz) [25].

| VWC (%) | 0 | 5 | 10 | 15 | 20 | 25 | 30 |
|---|---|---|---|---|---|---|---|
| $\varepsilon'_m$ | 3.7 | 4.01 | 5.29 | 7.22 | 9.78 | 13.4 | 16.7 |

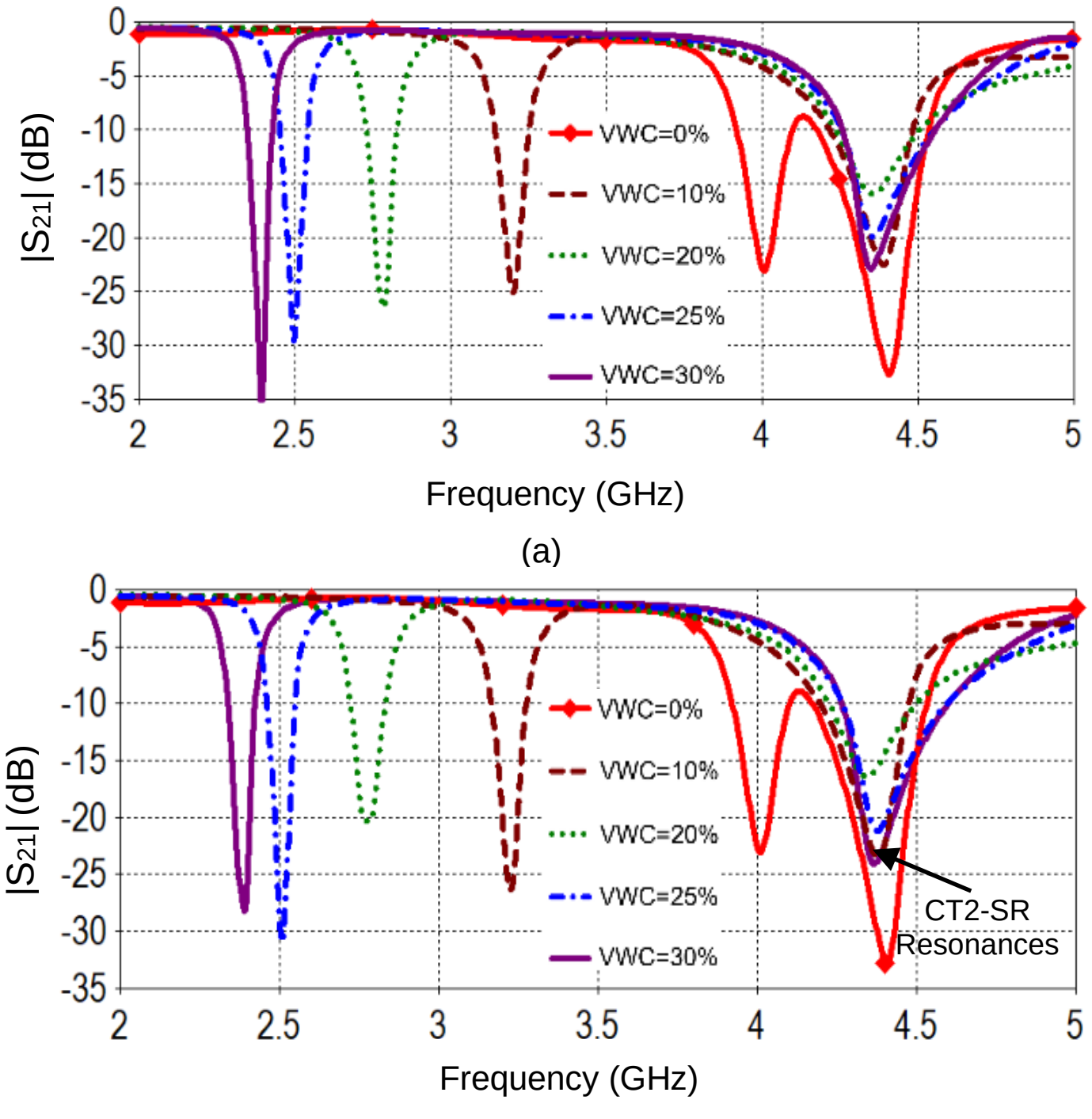

(b)

Fig. 5. |S21| results of the proposed DSMS for different VWC, a) Simulation, b) Measurement.

The measured difference between two zero transmissions ($f_d = f_{z2} - f_{z1}$) versus VWC (or relative permittivity) of the loaded sample (MUT) are plotted in Fig. 6(a). According to this figure, $f_d$ changes from 0.401 to 2.02 GHz for VWC range of 0% to 30%. By measuring $f_d$ between two resonance frequencies, the VWC value is calculated. In this regard, the measured values of $f_d$ are used to develop a mathematical model for the proposed sensor. To address this, a mathematical equasion is derived relating the frequency shift to VWC of the sandy soil moisture. The nonlinear least square curve fitting in MATLAB is used to derive two equations (5$^{th}$ and 7$^{th}$ degrees polynomial curve fitting method), describing the relation between $f_d$ variations as a function of VWC. Moreover, the error estimation of this mathematical modeling is approximately less than 18 MHz which is presented in Fig. 6(b).

Figure 7 demonstrates the sensitivity of resonance frequencies versus VWC for both resonators ($S_{fz1}, S_{fz2}$). According to this figure, $S_{fz1} = 110.8$ MHz and $S_{fz22} = 2.4$ MHz at VWC equal to 30% which verifies the DSMS performance as a differential sensor. This means, increasing the soil VWC pulls down $f_{z1}$ frequency, whereas the other resonance frequency, $f_{z2}$, is approximately fixed.

Figure 8 shows the effect of MUT thickness on the simulated resonance frequencies for $\varepsilon_{rm} = 16.7$. As can be seen, by increasing the MUT thickness to more than 30 mm ($0.6\lambda_g$ at 3.5 GHz), the resonance frequencies remain nearly constant. Hence, the sensor operation is independent of MUT thickness. This test can be easily performed by burying the sensor under a mass of soil.

Further, recently reported permittivity measurement sensors, [13]-[15], [26]-[31], are thoroughly compared in Table IV. These references are frequency domain techniques that mostly used SRR or CSRR configurations as the resonators in their structures. In Table IV, to compare given sensors at different operating frequencies, the fractional sensitivity is defined as:

$$Fractional\ Sensitivity\ (\%) = \frac{\Delta f}{f_{us}} \times 100 \qquad (6)$$

Where $\Delta f$ is the frequency shift for $\Delta\varepsilon_m = 1$ in the worst-case scenario (highest permittivity value of MUT), and $f_{us}$ is resonance frequency of unloaded sensor.

According to the results and Table IV, the proposed DSMS exhibits desirable sensitivity in a broad range of permittivity (1 to 16.5) in comparison with other works. For instance, although [26] and [27] achieved 130 MHz and 536 MHz frequency shifts in the permittivity measurement, their maximum permittivity values are 8 and 5 which are less than our sensor (16.5). Moreover, there are three differential sensors in Table IV while their maximum measurable permittivity is less than our proposed sensor. This comparison proves the usefulness of our proposed differential sensor technique for precision farming applications.

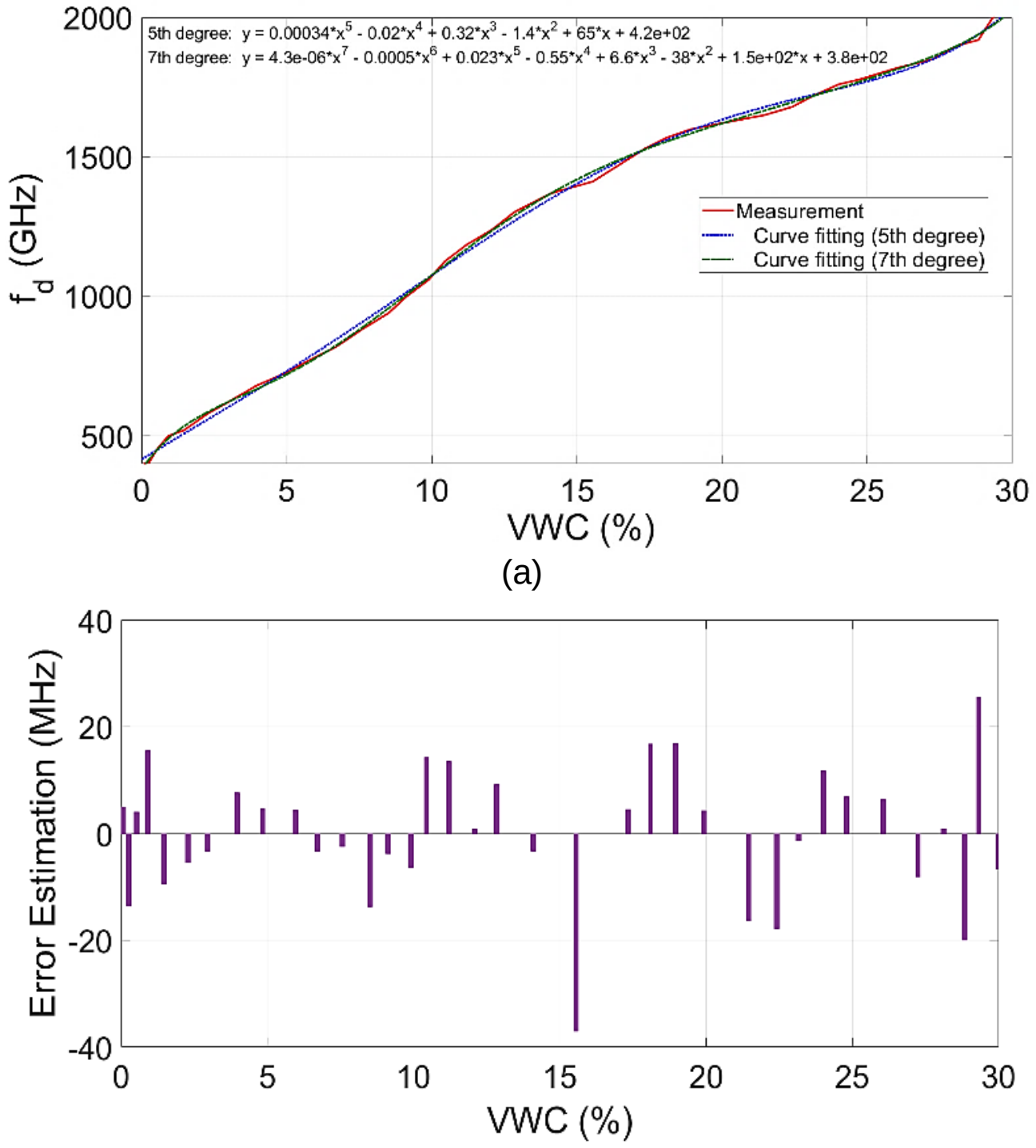

(b)

Fig. 6. a) Measured difference frequency ($f_d = f_{z2} - f_{z1}$), b) error estimation of the mathematical modeling, vs VWC for 5 trails.

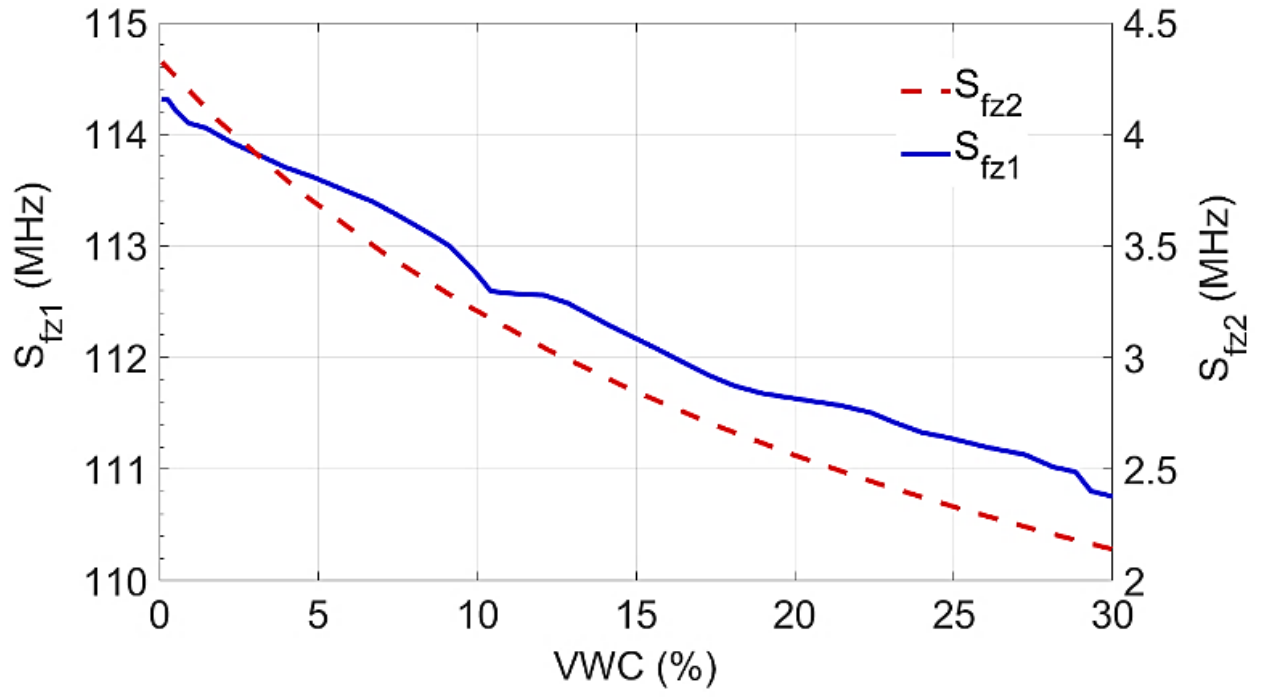

Fig. 7. Derivative of $f_d$ vs VWC in the proposed DSMS (sensor sensitivity).

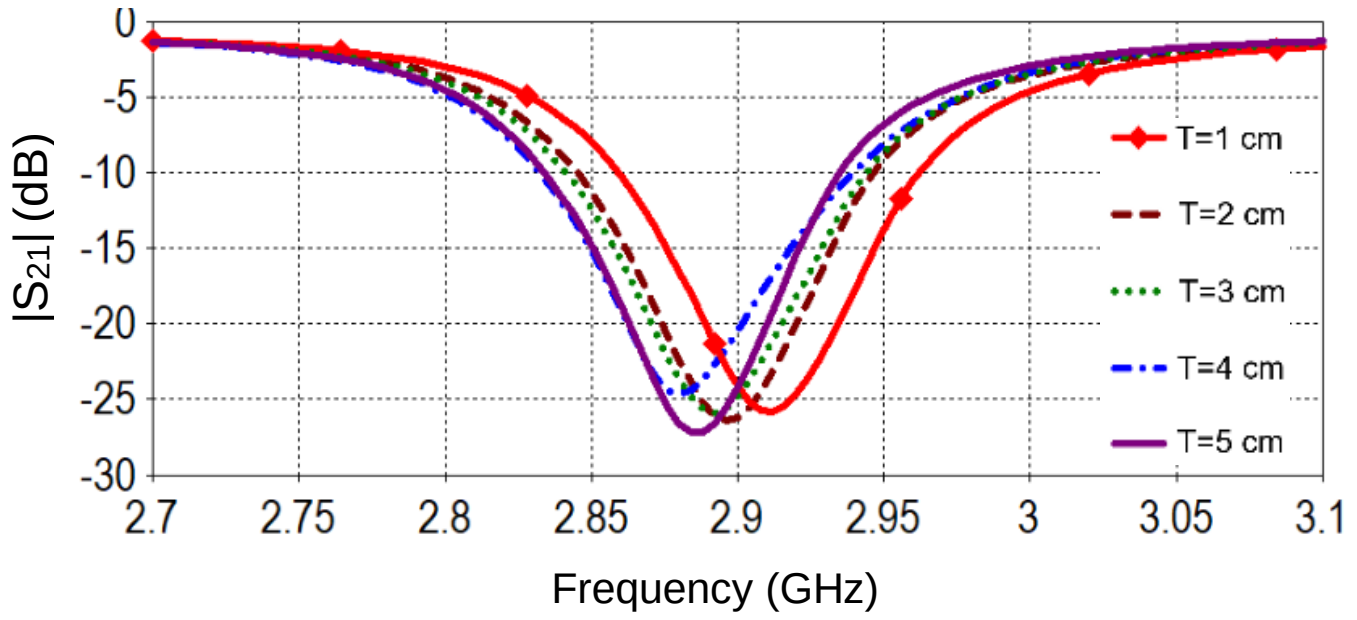

(a)

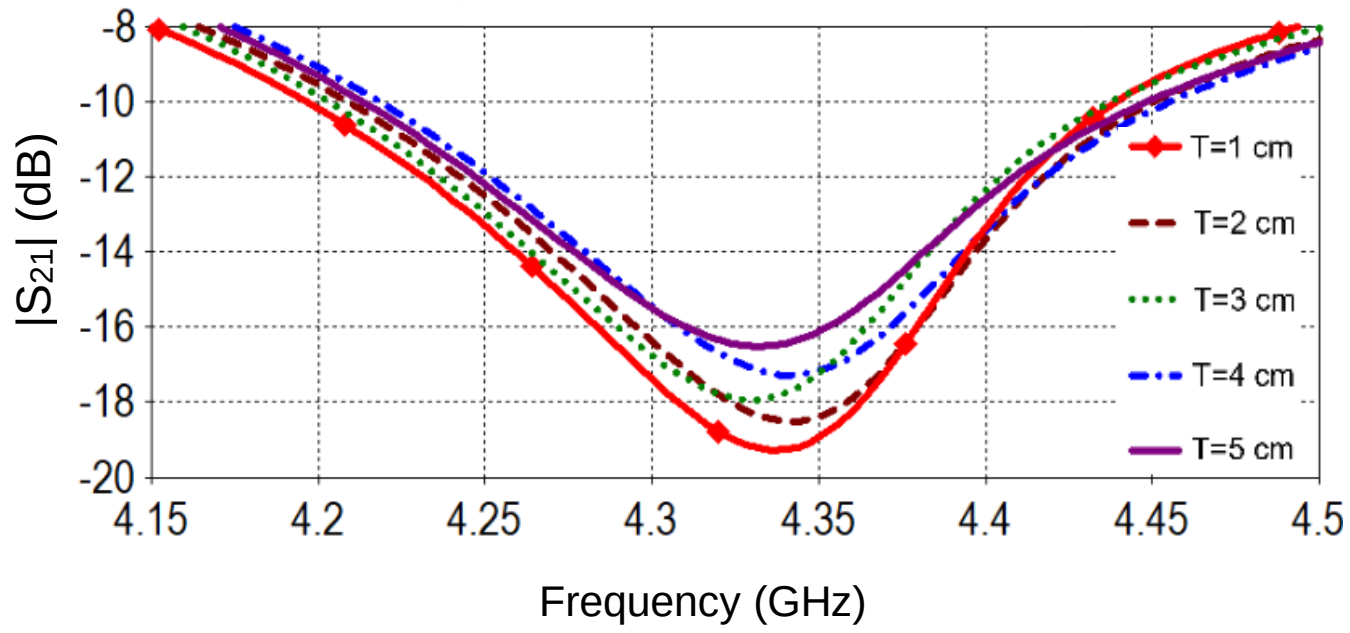

(b)

Fig. 8. Simulated $S_{21}$ of the sensor for different thicknesses (*T*) of the sample dielectric slab (1 cm<T<5 cm), a) T2-SR, b) CR2-SR.

## IV. Conclusion

In this paper, a new sensing device using uncoupled resonators is presented to realize a differential soil moisture sensor (DSMS). The proposed DSMS consists of a conventional microstrip line which is loaded with two resonators: T2-SR and CR2-SR. Simulated and measured DSMS exhibit sharp resonances which significantly improves the sensor sensitivity. The length and width of the proposed structure are approximately 10 mm and 50 mm, respectively. The low cost and miniaturized size of the proposed DSMS are the result of utilizing both sides of PCB (top and bottom layers) to realized resonators. Based on the theoretical analysis, simulation and achieved measurement results, this work proves the differential microwave sensing concept for precision agriculture. Moreover, the proposed technique can be used in the permittivity measurement of materials for other applications.

Table IV. Comparison table

| Ref. | $f_{us}$ (GHz) | Measurement Technique | Frequency Shift (Δf) for $\Delta\varepsilon_m = 1$ at max ($\varepsilon_m$) (MHz) | Fractional Sensitivity ($\frac{\Delta f}{f_{us}} \times 100$) (%) | Max. Measured Permittivity |
|---|---|---|---|---|---|
| [28] | 1.7 | SRR (Differential) | 33.3 | 1.9 | 10.2 |
| [27] | 6.1 | Stepped Impedance Resonators (Differential) | 536 | 8.8 | 5.7 |
| [15] | 2.4 | Coupled resonators | 13.4 | 0.56 | 30 |
| [13] | 2.1 | SRR (Differential) | 31 | 1.5 | 10.7 |
| [14] | 1.91 | SIR | 40 | 2.1 | 80 |
| [29] | 2.7 | CSRR | --- | --- | 10.2 |
| [30] | 4.5 | CSRR | 60 | 1.33 | NA |
| [26] | 4.5 | SRR | 130 | 2.8 | 8 |
| [31] | 5.5 | CSRR | 3.34 | 0.1 | 30 |
| This work | 4.5 | T2-SR and CR2-SR (Differential) | 110.8 MHz @ $\varepsilon'_{r1} = 16.7$ (VWC: 30%) | 2.5 | 16.7 |

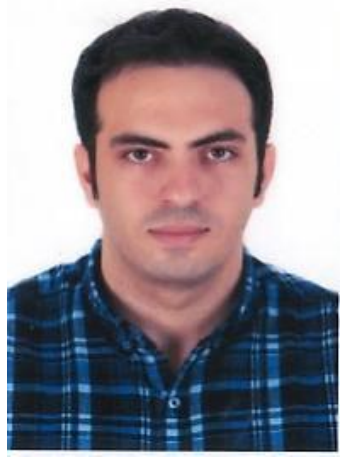

**Rasool Keshavarz** was born in Shiraz, Iran in 1986. He received the PhD degree in telecommunications engineering from the Amirkabir University of Technology, Tehran, Iran in 2017 and is currently working as Postdoctoral Research Associate in RFCT Lab at the University of Technology, Sydney, Australia. His main research interests are RF and microwave circuit and system design, sensors, antenna, digital Metamaterials, wireless power transfer (WPT) and RF energy harvesting (EH).

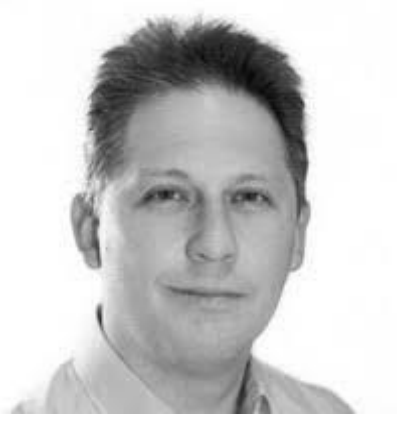

**Justin Lipman** (S'94, M'04, SM'12) received a PhD in Telecommunications Engineering from University of Wollongong, Australia in 2004. He is an Industry Associate Professor at the University of Technology Sydney (UTS) and a visiting Associate Professor at Hokkaido University's Graduate School of Engineering. Dr. Lipman is the Director of the RF Communications Technologies (RFCT) Lab, where he leads industry engagement in RF technologies, Internet of Things, Tactile Internet, Software Defined Communication and Agriculture 4.0. He serves as committee member in Standards Australia contributing to International IoT standards. Prior to joining UTS, Dr. Lipman was based in Shanghai, China and held a number of senior management and technical leadership roles at Intel and Alcatel driving research and innovation, product development, architecture and IP generation. He is an IEEE Senior Member. His research interests are in all "things" adaptive, connected, distributed and ubiquitous.

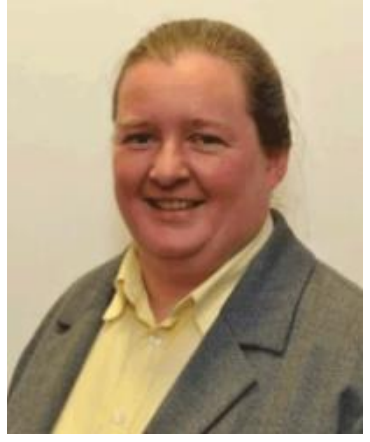

**Dominique M. M.-P. Schreurs** (Fellow, IEEE) received the M.Sc. degree in electronic engineering and the Ph.D. degree from the University of Leuven (KU Leuven), Leuven, Belgium, in 1992 and 1997, respectively. She has been a Visiting Scientist with Agilent Technologies, Santa Rosa, CA, USA, ETH Zürich, Zürich, Switzerland, and the National Institute of Standards and Technology, Boulder, CO, USA. She is currently a Full Professor with KU Leuven, where she is also the Chair of the Leuven ICT (the Leuven Centre on Information and Communication Technology). Her current research interests include the microwave and millimeter-wave characterization and modeling of transistors, nonlinear circuits, and bioliquids, and system design for wireless communications and biomedical applications. Prof. Schreurs served as the President of the IEEE Microwave Theory and Techniques Society from 2018 to 2019. She was an IEEE MTT-S Distinguished Microwave Lecturer. She has also served as the General Chair for the Spring Automatic RF Techniques Group (ARFTG) conferences in 2007, 2012, and 2018, and the President of the ARFTG organization from 2018 to 2019. She currently serves as the TPC Chair for the European Microwave Conference and also the Conference Co-Chair for the IEEE International Microwave Biomedical Conference. She was the Editor-in-Chief of the IEEE Transactions on Microwave Theory and Techniques.

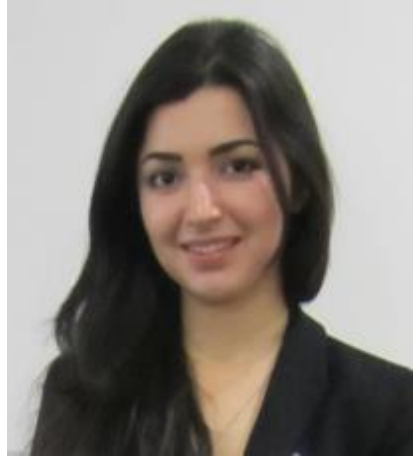

**Negin Shariati** is a Senior Lecturer in the School of Electrical and Data Engineering, Faculty of Engineering and IT, University of Technology Sydney (UTS), Australia. She established the state-of-the-art RF and Communication Technologies (RFCT) research laboratory at UTS in 2018, where she is currently the Co-Director and leads research and development in RF-Electronics, Sustainable Sensing, Low-power Internet of Things, and Energy Harvesting. She leads the Sensing Innovations Constellation at Food Agility CRC (Corporative Research Centre), enabling new innovations in agriculture technologies by focusing on three key interrelated streams; Energy, Sensing and Connectivity.

Since 2018, she has held a joint appointment as a Senior Lecturer at Hokkaido University, externally engaging with research and teaching activities in Japan.

She attracted over $650K worth of research funding over the past 3 years and across a number of CRC and industry projects, where she has taken the lead CI role and also contributed as a member of the CI team.

Negin Shariati completed her PhD in Electrical-Electronic and Communication Technologies at Royal Melbourne Institute of Technology (RMIT), Australia, in 2016. She worked in industry as an Electrical-Electronic Engineer from 2009-2012. Her research interests are in Microwave Circuits and Systems, RF Energy Harvesting, low-power IoT, Simultaneous Wireless Information and Power Transfer, AgTech, and Renewable Energy Systems.